\documentclass[preprint,aps,tightenlines,draft]{revtex4}
\usepackage{amsmath,amssymb}

\begin{document}
\def\eqn#1{Eq.$\,$#1}
\def\mb#1{\setbox0=\hbox{$#1$}\kern-.025em\copy0\kern-\wd0
\kern-0.05em\copy0\kern-\wd0\kern-.025em\raise.0233em\box0}
%\draft
\preprint{}

\title{Kinetic theory with angle-action variables}
\author{Pierre-Henri Chavanis}
\affiliation{Laboratoire de Physique Th\'eorique (UMR 5152 du CNRS), Universit\'e
Paul Sabatier\\ 118 route de Narbonne, 31062 Toulouse Cedex 4,
France\\ (chavanis{@}irsamc.ups-tlse.fr)}
%\date{}
%\maketitle

\begin{abstract}

\vskip 0.5cm

We present a kinetic theory for one-dimensional inhomogeneous systems
with weak long-range interactions. Starting from the Klimontovich
equation and using a quasilinear theory valid at order $1/N$ in a
proper thermodynamic limit $N\rightarrow +\infty$, we obtain a closed
kinetic equation describing the relaxation of the distribution
function of the system as a whole due to resonances between different
orbits. This equation is written in angle-action variables. It
conserves mass and energy and increases the Boltzmann entropy
(H-theorem). Using a thermal bath approximation, we derive a
Fokker-Planck equation describing the relaxation of a test particle
towards the Boltzmann distribution under the combined effects of
diffusion and friction.  We mention some analogies with the kinetic
theory of point vortices in two-dimensional hydrodynamics. We also
stress the limitations of our approach and the connection with recent
works.

\pacs{???}
\vskip0.5cm
\noindent Keywords: kinetic theory; long-range interactions; Fokker-Planck equations
\vskip0.5cm
\noindent Corresponding author:
P.H. Chavanis; e-mail: chavanis@irsamc.ups-tlse.fr; Tel:
+33-5-61558231; Fax: +33-5-61556065

\end{abstract}

\maketitle

\newpage

\section{Introduction}
\label{sec_introduction}

Systems with long-range interactions have been the object of considerable
interest in recent years \cite{dauxois}. Their dynamics is very rich and
presents many interesting features \cite{pa1}. Therefore, the
construction of a kinetic theory adapted to such systems is
valuable. Several kinetic theories have been developed in the
past. The first kinetic theory was constructed by Boltzmann
\cite{boltzmann} in his theory of gases. In that case, the particles do not
interact except during strong collisions. The extension of this
kinetic theory to the case of particles in Coulombian or Newtonian
interaction was considered by Landau \cite{landau} in the case of
plasmas and by Chandrasekhar \cite{chandra} in the case of stellar
systems. Developements and improvements of the kinetic theory of
Coulombian plasmas were made by Lenard \cite{lenard} and Balescu
\cite{balescu} using more formal approaches allowing to take into
account collective effects. They showed in particular how collective
effects can regularize the logarithmic divergence of the diffusion
coefficient at the Debye length. On the other hand, a kinetic theory
was developed by Dubin \& O'Neil \cite{dn,dubin} and Chavanis
\cite{preR,kin} to describe the ``collisional'' evolution of charged
rods in a magnetized plasma or point vortices in two-dimensional
hydrodynamics. In that case, the particles interact via a logarithmic
potential in two dimensions. A specificity of this model is that the
particles have no inertia so that the collision term acts in position
space. More recently, Bouchet, Dauxois and Chavanis [12-16]
 have considered the kinetic theory of the
Hamiltonian Mean Field (HMF) model, a toy model of systems with
long-range interactions which possesses a lot of interesting
properties and which can be studied in great detail. Finally, in very
recent works, Chavanis \cite{pa1,landaud} and Benedetti {\it et al.} 
\cite{benedetti} have developed the kinetic theory of a 2D Coulombian
plasma and Valageas \cite{valageas} has developed the kinetic theory
of a 1D gravitational system in a cosmological context.

Kinetic theories are important to describe the relaxation of a
Hamiltonian system of particles in interaction towards statistical
equilibrium. They are also necessary to determine the timescale of the
collisional relaxation and to prove whether the system will truly
relax towards statistical equilibrium. Indeed, the convergence towards
statistical equilibrium is not firmly established for complex systems
with long-range interactions.  A first reason is that systems with
long-range interactions may exhibit non-markovian effects and spatial
delocalization so that the Boltzmann $H$-theorem is difficult to prove
and could even be wrong in a strict sense. Such problems have been
discussed for example by Kandrup [20-22] in
the case of self-gravitating systems and by Chavanis for point
vortices \cite{kin} and for the HMF model \cite{curious}. These
authors derive {\it exact} kinetic equations for these systems which
apparently do not satisfy an $H$-theorem. The $H$-theorem is recovered
only when additional assumptions (markovian approximation, local
approximation, neglect of three body correlations,...) are implemented
\footnote{These assumptions can be justified in a proper thermodynamic
limit with $N\rightarrow +\infty$. Therefore, the $H$-theorem holds in
that limit. This is consistent with the results of equilibrium statistical
mechanics which indicate that, for $N\rightarrow +\infty$, the
distribution function $f({\bf r},{\bf v})$ of statistical equilibrium
 maximizes the Boltzmann entropy $S_{B}[f]$ at fixed mass $M[f]$
and energy $E[f]$; see \cite{pa1} for more details and
references.}. On the other hand, the relaxation of the system is due
to a condition of resonance between different orbits and it may happen
that the relaxation stops because there is no resonance anymore.  This
is the case for point vortices in two-dimensional hydrodynamics when
the profile of angular velocity becomes monotonic
\cite{dn,kin}. This is also the case for spatially
homogeneous 1D systems for which the Landau-Lenard-Balescu collision
term cancels out \cite{bd,cvb,pa1}. In these cases, the system remains
frozen in a long-lived ``metastable'' state (a stationary solution of
the Vlasov equation) which is not the statistical equilibrium
state. This means either that the system will not reach statistical
equilibrium or that the kinetic theory is incomplete.  Indeed, the
collision term calculated in the above-mentioned kinetic theories
corresponds to a term of order $1/N$ in a systematic expansion carried
out in a proper thermodynamic limit $N\rightarrow +\infty$ (one can
rescale the parameters so that the coupling constant scales like
$k\sim 1/N$ while $E\sim N$, $T\sim 1$ and $V\sim 1$
\cite{pa1}). Therefore, they describe the dynamics of the system on a
timescale of order $Nt_{D}$ where $t_{D}$ is the dynamical time. If
the collision term cancels out because of the absence of resonances,
the relaxation towards statistical equilibrium (if any) will take
place on a timescale larger than $Nt_{D}$ so that next order terms in
the $1/N$ expansion must be taken into account. The calculation of
these higher order terms has not been realized for any of the kinetic
theories mentioned previously. However, for the HMF model, it is shown
numerically that the relaxation towards statistical equilibrium takes
place on a timescale of order $N^{1.7}t_{D}$
\cite{yamaguchi}.

In all these works (except for point vortices), the system is assumed
to be spatially homogeneous \cite{landaud}. For the HMF model, a
homogeneous phase exists if the energy is larger than a critical
energy $E_{c}$. For Coulombian plasmas, the interaction is shielded on
a distance corresponding to the Debye length so that the system is
spatially homogeneous. Stellar systems, on the other hand, are clearly
inhomogeneous but most kinetic theories assume that the encounters can
be treated with a {\it local approximation} \cite{bt} as if the system
were spatially homogeneous. This results in a logarithmic
divergence of the diffusion coefficient which is cured by introducing
an appropriate cut-off at the Jeans scale. It is expected that this
divergence would be removed if effects of spatial inhomogeneity were
taken into account. Therefore, the description of the dynamics of
inhomogeneous systems with long-range interactions, like the clustered
phase of the HMF model below the critical energy $E_{c}$, remains a
challenging problem. In a previous paper \cite{curious}, we have given
some elements of kinetic theory for the inhomogeneous HMF model in a
very special situation of low energy (close to the ground state) where
the particles are strongly concentrated around a cluster at $\theta=0$
so that they all perform a harmonic motion with the {\it same}
frequency. In that ``pathological'' case, some curious behaviors have
been predicted. However, in more general situations, the particles
will have different frequencies and we may expect that the relaxation
will be due to some sort of resonances as in the case of 2D point
vortices
\cite{dn,kin}.

In this paper, we present a kinetic theory for one-dimensional
inhomogeneous systems with weak long-range interactions.  Starting
from the Klimontovich equation and using a quasilinear theory
(Sec. \ref{sec_klim}) valid at order $1/N$ in a proper thermodynamic
limit $N\rightarrow +\infty$, we obtain a closed kinetic equation
describing the relaxation of the distribution function of the system
as a whole due to resonances between different orbits
(Sec. \ref{sec_q}). This equation is written in angle-action variables
(Sec. \ref{sec_aa}). It conserves mass and energy and increases the
Boltzmann entropy so that a H-theorem can be proved
(Sec. \ref{sec_p}). Using a thermal bath approximation
(Sec. \ref{sec_tb}), we derive a Fokker-Planck equation that describes
the relaxation of a test particle towards the Boltzmann distribution
under the combined effects of diffusion and friction.  Unfortunately,
our derivation of the kinetic equation is based on a factorization
hypothesis (\ref{q7}) that we cannot justify and that is probably not
rigorously valid. However, we expect that more general approaches will
be able to circumvent this difficulty. The important point is that the
final kinetic equation (\ref{q16}) that we obtain is closed, without
undetermined parameter, and that it possesses a lot of interesting
properties (conservation laws and H-theorem). This gives some
confidence in its relevance to describe the dynamics of the system
even if the factorization hypothesis does not hold. In fact, in a very
recent work, Valageas \cite{valageas} has independently tackled a
similar problem with a different method that avoids the factorization
hypothesis. We shall discuss the connection between the two approaches
in Sec. \ref{sec_conn} and show that our kinetic equation (\ref{q16})
can also be obtained from the approach of Valageas
\cite{valageas}. This provides therefore a rigorous justification 
for this kinetic equation without {\it ad hoc} assumption.

\section{The Klimontovich equation} \label{sec_klim}

Let us consider a one-dimensional Hamiltonian system of particles in
interaction described by the equations of motion
\begin{eqnarray}
\label{klim1}
\frac{dx_{i}}{dt}=\frac{\partial H}{\partial v_{i}}, \qquad \frac{dv_{i}}{dt}=-\frac{\partial H}{\partial x_{i}},\nonumber\\
{H}=\frac{1}{2}\sum_{i=1}^{N}v_{i}^{2}+\sum_{i<j}u(x_{i}-x_{j}).
\end{eqnarray}
We assume that the particles interact via a weak long-range binary
potential $u=u(|x_{i}-x_{j}|)$. This assumption is necessary to
justify the quasilinear theory that we shall develop in the
sequel. This potential of interaction can be, for example,
the potential $u=-(k/2\pi)\cos(\theta_{i}-\theta_{j})$ of the HMF
model \cite{cvb}. We introduce the exact distribution function
$f_{d}(x,v,t)=\sum_{i=1}^{N}
\delta(x-x_{i}(t))\delta(v-v_{i}(t))$ and the corresponding 
potential $\Phi_{d}(x,t)=\int u(x-x')f_{d}(x',v',t)dx'dv'$. The exact distribution function satisfies the Klimontovich equation
\begin{equation}
\label{klim2}
\frac{\partial f_{d}}{\partial t}+\lbrack H_{d},f_{d}\rbrack=0,
\end{equation}
where $H_{d}=v^{2}/2+\Phi_{d}(x,t)$ is the exact energy of an
individual particle and we have introduced the Poisson bracket
\begin{equation}\label{klim3}
\lbrack H,f\rbrack=\frac{\partial H}{\partial {p}}\frac{\partial f}{\partial {q}}-\frac{\partial H}{\partial {q}}\frac{\partial f}{\partial {p}},
\end{equation}
with $H=H_{d}$, $f=f_{d}$, ${q}={x}$ and ${p}={v}$. The Klimontovich
equation (\ref{klim2}) contains exactly the same information as the Hamilton
equations (\ref{klim1}).  

We introduce a smooth distribution function $f=\langle f_{d}\rangle$
which is the statistical average of the exact distribution
function. We note $\Phi=\langle \Phi_{d}\rangle$ the corresponding
potential.  Then, we write $f_{d}=f+\delta f$ and
$\Phi_{d}=\Phi+\delta\Phi$. Inserting this decomposition in the
Klimontovich equation (\ref{klim2}), we obtain
\begin{equation}\label{klim4}
\frac{\partial f}{\partial t}+\frac{\partial \delta f}{\partial t}+\lbrack H,f\rbrack+\lbrack H,\delta f\rbrack+\lbrack \delta\Phi, f\rbrack+\lbrack \delta\Phi, \delta f\rbrack =0,
\end{equation}
where we have used $H_{d}=H+\delta H$ with $H=v^{2}/2+\Phi(x,t)$ and
$\delta H=\delta\Phi$.  Taking the average of Eq. (\ref{klim4}), we get
\begin{equation}\label{klim4b}
\frac{\partial f}{\partial t}+\lbrack H,f\rbrack=-\langle \lbrack \delta\Phi, \delta f\rbrack \rangle.
\end{equation}
Subtracting this relation from Eq. (\ref{klim4}) and neglecting nonlinear terms, we obtain an equation for the perturbation
\begin{equation}\label{klim5}
\frac{\partial \delta f}{\partial t}+\lbrack H,\delta f\rbrack+\lbrack \delta\Phi, f\rbrack=0.
\end{equation}
This linearization is the essence of the quasilinear theory
\cite{pitaevskii}.  This approximation can be shown to be equivalent
to keeping only terms of order $1/N$ in a proper thermodynamic limit
$N\rightarrow +\infty$ \cite{pa1}. In the $N\rightarrow +\infty$
limit, the term in the right hand side of Eq. (\ref{klim4b}) vanishes and we
obtain the Vlasov equation
\begin{equation}\label{klim6}
\frac{\partial f}{\partial t}+\lbrack H,f\rbrack=0.
\end{equation}
This equation admits an infinite number of stationary solutions of the
form $f=f(\epsilon)$ where $\epsilon=v^{2}/2+\Phi(x)$ is the
individual energy. Such steady distributions can be reached through a
process of violent relaxation involving only meanfield effects
\cite{lb,next5}. If we keep terms of order $1/N$ or smaller, we shall 
obtain a kinetic equation taking into account the developement of
correlations between particles due to granularities (finite $N$
effects). This is similar to the effect of ``collisions'' in the
theory of gases. Thus, the right hand side of Eq. (\ref{klim4b}) is
similar to the collision term in the Boltzmann equation.  This term is
expected to drive the system towards statistical
equilibrium. Therefore, it is expected to select the Boltzmann
distribution among all possible steady solutions of the Vlasov
equation (see, however, the discussion of Sec. \ref{sec_eq}).  Since
this term is of order $1/N$ or smaller, the effect of ``collisions'' is
a slow process that takes place on a timescale $Nt_{D}$ or
larger. Therefore, there is a scale separation between the dynamical
time $t_{D}$ which is the timescale at which the system reaches a
steady state of the Vlasov equation through violent relaxation due to
inertial effects and the ``collisional'' time $t_{coll}$ at which the
system is expected to relax towards statistical equilibrium due to
thermodynamical effects. Because of this scale separation, the
distribution function will pass by a series of stationary solutions of
the Vlasov equation depending only on energy, slowly evolving with time due to collisions
(finite $N$ effects). Hence, the distribution function can be
approached by
\begin{equation}\label{klim7}
f(x,v,t)\simeq f(\epsilon,t).
\end{equation}
This means that the system is approximately in mechanical equilibrium
\footnote{It is easy to show that if the distribution function $f=f(\epsilon,t)$ depends only 
on the energy, then the condition of hydrostatic equilibrium $\nabla
p+\rho\nabla\Phi={\bf 0}$ holds at any time; see, e.g., Eq. (33) in
\cite{virial}.} at each stage of the dynamics (due to violent
relaxation) and that the ``collisions'' will drive it to
thermodynamical equilibrium. The purpose of this paper is to obtain a
closed expression for the collision term in energy space when the
rapid dynamics has been averaged over the orbits. This is similar to
the orbit-averaged-Fokker-Planck equation in astrophysics \cite{bt}.

\section{Angle-action variables} \label{sec_aa}

The previous decomposition has separated the smooth meanfield
distribution $f(\epsilon,t)$ which evolves slowly with time from the
fluctuations $\delta f$ which are expected to have a decorrelation
time on a much shorter timescale \footnote{This assumption is not
completely obvious for systems with long-range interactions. For
example, in the case of 3D stellar systems, the correlations of the
force decrease algebraically, like $t^{-1}$ \cite{chandrat}. For the
one-dimensional HMF model at $T=0$, the force auto-correlation
function has an oscillatory behaviour \cite{curious}. For larger
values of $T$, one expects an exponential decay of the force
auto-correlation function as in the homogeneous case
$T>T_{c}$. However, the timescale of the exponential relaxation can be
very large, especially close to the critical point $T_{c}$
\cite{bouchet,cvb}. In
any case, in the $N\rightarrow +\infty$ limit, there will be a
time-scale separation between the decorrelation time of the
fluctuations and the evolution of the smooth meanfield distribution
which justifies the decomposition made in this paper.  }. For
inhomogeneous systems, it proves convenient to introduce angle-action
variables \cite{bt} constructed from the slowly evolving
distribution. Then, when we focus on the evolution of the
perturbation, we can consider that the smooth distribution is
``frozen''. This will allow us to determine the collision term. Then,
we can take into account the time dependence of the smooth
distribution function through the kinetic equation (\ref{klim4b}). We
define the action
\begin{equation}\label{aa1}
J=\frac{1}{2\pi}\oint v dx,
\end{equation}
where
\begin{equation}\label{aa2}
v(x)=\sqrt{2(\epsilon-\Phi(x))}.
\end{equation}
By definition, the action is a function $J=J(\epsilon)$ of the energy
so it is constant on the iso-energetic lines in phase space. As
discussed previously, we assume in a first step that the potential
$\Phi(x,t)$ is ``frozen'' on the timescale on which the fluctuations
have essential correlations so that $J$ depends only on the energy (it
is a new coordinate). We also introduce an angle variable
\begin{equation}\label{aa3}
\phi=\Omega\int^{x}\frac{dx}{v}.
\end{equation}
This angle parameterizes the iso-energetic line with energy
$\epsilon$. Therefore, a particle with coordinates $(x,v)$ in phase-space can 
be described equivalently  by the angle-action variables $(\phi,J)$.
Noting that the period of an orbit with energy $\epsilon$ is
\begin{equation}\label{aa4}
T(\epsilon)=\oint dt=\oint \frac{dx}{v}=\frac{2\pi}{\Omega(\epsilon)},
\end{equation}
we find that $\Omega=\Omega(\epsilon)=\Omega(J)$ is the pulsation of the orbit with energy $\epsilon$. We also note for future reference that $J(\epsilon)$ represents the surface of phase space with energy smaller than $\epsilon$. Therefore, the surface of phase space with energy between $\epsilon$ and $\epsilon+d\epsilon$ is $g(\epsilon)d\epsilon$ with
\begin{equation}\label{aa5}
g(\epsilon)=\frac{dJ}{d\epsilon}.
\end{equation}
This is similar to the density of states in statistical mechanics (but
here in the $\mu$-space). Now, from Eq. (\ref{aa1}), we have
\begin{equation}\label{aa6}
\frac{d J}{d\epsilon}=\frac{1}{2\pi}\oint \frac{dx}{\sqrt{2(\epsilon-\Phi)}}=\frac{1}{2\pi}\oint \frac{dx}{v}=\frac{1}{\Omega(\epsilon)},
\end{equation}
so that $g(\epsilon)=1/\Omega(\epsilon)$. Accordingly, the density  of particles with energy $\epsilon$ is
\begin{equation}\label{aa7}
N(\epsilon)=f(\epsilon)g(\epsilon)=\frac{f(\epsilon)}{\Omega(\epsilon)}.
\end{equation}

Since only the meanfield Hamiltonian appears in the advective term of
Eqs. (\ref{klim4b})-(\ref{klim5}), it only involves the unperturbed
motion of the particles. The smooth Hamiltonian equations for the
conjugate variables $(q,p)$ can be written
\begin{equation}\label{aa8}
\frac{dq}{dt}=\frac{\partial H}{\partial p}, \qquad \frac{dp}{dt}=-\frac{\partial H}{\partial q}.
\end{equation}
If we use the variables $(x,v)$, we find that the dynamics is
relatively complicated because the potential explicitly appears in the
second equation. Therefore, this equation $dv/dt=-\Phi'(x)$ cannot
be easily integrated except if $\Phi=0$, i.e. for a homogeneous
system. In that case, the unperturbed equations of motion reduce to
$x=vt+x_{0}$, i.e. a motion at constant velocity.  Now, the
angle-action variables are constructed so that the Hamiltonian does
not depend on the angles $\phi$. Therefore, the Hamilton equations for
the conjugate variables $(\phi,J)$ are
\begin{equation}
\label{aa9}
\frac{d\phi}{dt}=\frac{\partial H}{\partial J}=\Omega(J), \qquad \frac{dJ}{dt}=-\frac{\partial H}{\partial \phi}=0.
\end{equation}
From these equations, we find that $J={\rm Cst.}$ and
$\phi=\Omega(J)t+\phi_{0}$ so that the equations of motion are very
simple in such variables. They extend naturally the trajectories with
constant velocity for homogeneous systems. This is why this choice of
variables is relevant to develop the kinetic theory.  Now, in terms of
the angle-action variables, using the relations
\begin{equation}\label{aa10}
\lbrack H,\delta f\rbrack=\frac{\partial H}{\partial J}\frac{\partial \delta f}{\partial \phi}-\frac{\partial H}{\partial \phi}\frac{\partial \delta f}{\partial J}=\Omega(J)\frac{\partial \delta f}{\partial \phi},
\end{equation}
\begin{equation}\label{aa11}
\lbrack \delta \Phi,f\rbrack=\frac{\partial \delta\Phi}{\partial J}\frac{\partial f}{\partial \phi}-\frac{\partial \delta\Phi}{\partial \phi}\frac{\partial f}{\partial J}=-\frac{\partial \delta\Phi}{\partial \phi}\frac{\partial f}{\partial J},
\end{equation}
we find that Eqs.  (\ref{klim4b})-(\ref{klim5}) take the form
\begin{equation}\label{aa12}
\frac{\partial f}{\partial t}=\frac{\partial}{\partial J}\langle  \delta f \frac{\partial\delta\Phi}{\partial\phi}\rangle,
\end{equation}
and
\begin{equation}\label{aa13}
\frac{\partial\delta f}{\partial t}+\Omega(J)\frac{\partial\delta f}{\partial\phi}-\frac{\partial\delta\Phi}{\partial\phi}\frac{\partial f}{\partial J}=0.
\end{equation}

\section{The quasilinear theory} \label{sec_q}

The equations (\ref{aa12})-(\ref{aa13}) form the starting point of the
quasilinear theory.  The objective is to solve Eq. (\ref{aa13}) for
the fluctuation $\delta f$ and substitute the result back into
Eq. (\ref{aa12}). Here, we follow the approach of \cite{kp,paddy}. We
introduce the Fourier transforms \footnote{ The Fourier transforms on
$t$ assume that the perturbation behaves as $e^{-i\omega t}$ both at
$t\rightarrow -\infty$ and $t\rightarrow +\infty$. Since a real system
is set up at a fixed time $t=t_{0}$, we have an {\it initial value
problem} and it would be more relevant to use a Laplace transform with
respect to $t$, from $t=t_{0}$ to $t=\infty$ (the conventional Laplace
transform uses $-s$ in place of $-i\omega$ in the integrand); see
Lifshitz \& Pitaevskii
\cite{pitaevskii} and Binney \& Tremaine \cite{bt}, Appendix 5.A for
more details on these technical issues.}:
\begin{equation}\label{q1}
\delta f(\phi,J,t)=\sum_{m}\int \delta\hat{f}_{m}(J,\omega)e^{im\phi}e^{-i\omega t}d\omega,
\end{equation}
\begin{equation}\label{q2}
\delta \Phi(\phi,J,t)=\sum_{m}\int \delta\hat{\Phi}_{m}(J,\omega)e^{im\phi}e^{-i\omega t}d\omega.
\end{equation}
Substituting these expressions in Eq.  (\ref{aa13}), we obtain
\begin{equation}\label{q3}
\left\lbrack \omega-m\Omega(J)\right\rbrack \delta\hat{f}_{m}=-m\frac{\partial f}{\partial J}\delta\hat{\Phi}_{m}. 
\end{equation}
Therefore, the perturbed distribution function is given by
\begin{equation}
\label{q4}
\delta\hat{f}_{m}(J,\omega)=\frac{-m\frac{\partial f}{\partial J}(J)}{\omega-m\Omega(J)}
\delta\hat{\Phi}_{m}(J,\omega)+\hat{g}_{m}(J,\omega), 
\end{equation}
where  $\hat{g}_{m}(J,\omega)$ is  solution to the equation
\begin{equation}\label{q5}
\left\lbrack \omega-m\Omega(J)\right\rbrack \hat{g}_{m}(J,\omega)=0,
\end{equation}
and is related to the initial conditions \cite{pitaevskii,kp,paddy}. Here,
$\hat{g}_{m}(J,\omega)$ arises because of the discrete nature of the
system, i.e. the fact that the exact distribution $f_{d}$ is a sum of
$\delta$-functions.

The Fourier transforms
of the potential and of the distribution function are related to each
other by an equation of the form (Appendix \ref{sec_amn})
\begin{equation}\label{q6}
\delta\hat{\Phi}_{m}(J,\omega)=\sum_{m'}\int A_{mm'}(J,J')\delta\hat{f}_{m'}(J',\omega)dJ',
\end{equation}
where $A_{mm'}(J,J')$ are the Fourier components of the potential of
interaction written in angle-action variables.  At that stage, we
shall make a rough approximation and assume that the function
$A_{mm'}(J,J')$ can be factorized \footnote{The same approximation is
made by Pichon \cite{pichon} to investigate the linear dynamical
stability of a steady solution of the Vlasov equation (\ref{klim6}).} 
according to
\begin{equation}
\label{q7}
A_{mm'}(J,J')=A_{m}(J)A_{m'}(J').
\end{equation}
Although this approximation is clearly a weakness of our approach, we
can give the following arguments to motivate our study: 1. Without a
decomposition of this type, it appears difficult to go any further in
our approach. 2. There may exist some potentials of interaction for
which this approximation holds exactly or approximately. 3. Such
factorization could be obtained rigorously by using another system of
coordinates or another basis on which the potential of interaction
could be developed. 4. This approximation does not seem to be crucial:
in particular the final kinetic equation (\ref{q16}) can be expressed
in terms of $A_{mm'}(J,J')$ alone provided that we neglect collective
effects. Therefore, the approximation (\ref{q7}) may lead to an inaccurate
description of collective effects while preserving the correct
structure of the kinetic equation. Assuming that Eq. (\ref{q7}) holds, we
obtain
\begin{equation}
\label{q8}
\frac{\delta\hat{\Phi}_{m}(J,\omega)}{A_{m}(J)}=\sum_{m'}\int A_{m'}(J')\delta\hat{f}_{m'}(J',\omega)dJ'.
\end{equation}
This ratio  is independent on $m$ and $J$. Substituting Eq. (\ref{q4}) in Eq. (\ref{q8}) we find
\begin{equation}\label{q9}
\frac{\delta\hat{\Phi}_{m}(J,\omega)}{A_{m}(J)}=-\sum_{m'}\int A_{m'}(J')^{2}
\frac{m'\frac{\partial f}{\partial J}(J')}{\omega-m'\Omega(J')}
\frac{\delta\hat{\Phi}_{m'}(J',\omega)}{A_{m'}(J')}dJ'+\sum_{m'}\int A_{m'}(J')
\hat{g}_{m'}(J',\omega)dJ'.
\end{equation}
Using the remark following Eq. (\ref{q8}), this can be rewritten
\begin{eqnarray}\label{q10}
\frac{\delta\hat{\Phi}_{m}(J,\omega)}{A_{m}(J)}\left\lbrace 1+\sum_{m'}\int A_{m'}(J')^{2}
\frac{m'\frac{\partial f}{\partial J}(J')}{\omega-m'\Omega(J')}dJ'\right\rbrace=\sum_{m'}\int A_{m'}(J')
\hat{g}_{m'}(J',\omega)dJ'.
\end{eqnarray}
We introduce the dielectric function
\begin{eqnarray}\label{q11}
\epsilon(\omega)=1+\sum_{m'}\int A_{m'}(J')^{2}
\frac{m'\frac{\partial f}{\partial J}(J')}{\omega-m'\Omega(J')}dJ',
\end{eqnarray}
so that
\begin{equation}\label{q12}
\delta\hat{\Phi}_{m}(J,\omega)=\frac{A_{m}(J)}{\epsilon(\omega)}\sum_{m'}\int A_{m'}(J')
\hat{g}_{m'}(J',\omega)dJ'.
\end{equation}
At this stage, we must require that $\epsilon(\omega)\neq 0$ for any
{\it real} $\omega$. This is the case if the distribution function
$f(\epsilon)$ is stable with respect to the Vlasov equation (see
Appendix \ref{sec_dis}). In fact, the distribution function
$f(\epsilon)$ which depends only on the energy results from a
dynamical relaxation process as described at the end of
Sec. \ref{sec_klim}. This dynamical process (violent relaxation)
necessarily puts the system on a stable stationary solution of the Vlasov
equation. Therefore, $f(\epsilon,t)$ is a stable steady solution of the
Vlasov equation at each stage of the dynamics which slowly evolves
with time due to finite $N$ effects (collisions).

Substituting Eq. (\ref{q12}) in Eq. (\ref{q4}), we obtain
\begin{equation}\label{q13}
\delta\hat{f}_{m}(J,\omega)=\frac{-m\frac{\partial f}{\partial J}(J)}{\omega-m\Omega(J)}\frac{A_{m}(J)}{\epsilon(\omega)}\sum_{m'}\int A_{m'}(J')
\hat{g}_{m'}(J',\omega)dJ'+\hat{g}_{m}(J,\omega).
\end{equation}
We can now compute the diffusion current using
\begin{equation}\label{q14}
\langle \delta f \frac{\partial\delta\Phi}{\partial\phi}\rangle=i\sum_{m,m'}m'\int d\omega d\omega'\langle \delta\hat{f}_{m}(J,\omega) \delta\hat{\Phi}_{m'}(J,\omega')\rangle e^{i(m+m')\phi}e^{-i(\omega+\omega')t}, 
\end{equation}
and the expressions (\ref{q12}) and (\ref{q13}). From that point, the
calculations are similar to those carried out for the usual
quasilinear theory
\cite{pitaevskii,kp,paddy} and they will not be repeated (see, e.g., Appendix B of \cite{pa1}). We finally obtain the kinetic equation
\begin{eqnarray}\label{q15}
\frac{\partial f}{\partial t}=\frac{1}{2}\frac{\partial}{\partial J}\sum_{m,m'}\int \frac{1}{|\epsilon(m'\Omega(J'))|^{2}}m A_{m}(J)^{2}A_{m'}(J')^{2}\delta(m\Omega(J)-m'\Omega(J'))\nonumber\\
\times\left\lbrace f(J')m\frac{\partial f}{\partial J}-f(J)m'\frac{\partial f}{\partial J'}\right\rbrace dJ'.
\end{eqnarray}
Using Eq. (\ref{q7}), the product $A_{m}(J)A_{m'}(J')$ can be replaced by
$A_{mm'}(J,J')$. Then, we note that the function $A_{m}(J)$ occurs
only in the dielectric function. From now on, we shall neglect
collective effects and take $|\epsilon(m'\Omega(J'))|=1$. It is likely
that our approach does not describe collective effects accurately.
Therefore, the kinetic equation that we shall consider is
\begin{equation}
\label{q16}
\frac{\partial f}{\partial t}=\frac{1}{2}\frac{\partial}{\partial J}\sum_{m,m'}\int {m A_{mm'}(J,J')^{2}}\delta(m\Omega(J)-m'\Omega(J'))\left\lbrace f(J')m\frac{\partial f}{\partial J}-f(J)m'\frac{\partial f}{\partial J'}\right\rbrace dJ'.
\end{equation}
This equation depends only on $A_{mm'}(J,J')$ which can be calculated
from Eq. (\ref{amn5}) for any potential. It does not depend on the
individual terms $A_{m}(J)$ and $A_{m'}(J')$ which would be
undetermined if Eq. (\ref{q7}) were not valid. Therefore, the kinetic
equation (\ref{q16}) is well-posed for any potential.  This may be an
indication that its form does not crucially depend on the
approximations that we have made in the course of the
derivation. Furthermore, we shall see in the next section that this
kinetic equation respects all the conservation laws of the Hamiltonian
system (conservation of mass and energy) and increases the Boltzmann
entropy. This again suggests that this equation provides a good
description of the out-of-equilibrium dynamics even if our manner to
derive it is not entirely satisfactory.  This equation is the main
result of the paper.  It can be viewed as the counterpart of the
Landau-Lenard-Balescu equation of plasma physics for spatially
inhomogenous one-dimensional systems with weak long-range
interactions. We note, in particular, that the relaxation is due to a
condition of resonance $m\Omega(J)=m'\Omega(J')$ between different
orbits. This is similar to the kinetic equation obtained for point vortices in two-dimensional hydrodynamics \cite{kin}:
\begin{equation}
\label{q17}
\frac{\partial P}{\partial t}=-\frac{N\gamma^{2}}{4r}\frac{\partial}{\partial r}\int_{0}^{+\infty}r_{1}dr_{1}\delta (\Omega(r)-\Omega(r_{1}))\ln\left\lbrack 1-\left (\frac{r_{<}}{r_{>}}\right )^{2}\right\rbrack \left (\frac{1}{r}P_{1}\frac{\partial P}{\partial r}-\frac{1}{r_{1}}P\frac{\partial P_{1}}{\partial r_{1}}\right ),
\end{equation} 
where $\Omega(r)$ is the angular velocity, $\gamma$ is the circulation
of a point vortex and $r_{<}$ (resp. $r_{>}$) denotes the smallest
(resp. largest) value of $r$ and $r_{1}$ . Here again, the
relaxation is due to a condition of resonance
$\Omega(r)=\Omega(r_{1})$ when the profile of angular velocity is
non-monotonic.  Equation (\ref{q17}) also neglects collective effects
but the more accurate treatment of Dubin \& O'Neil \cite{dn} allows to
take them into account.

\section{Properties of the kinetic equation} \label{sec_p}

Let us derive some general properties of the kinetic equation (\ref{q16}). This equation can be written in the form of a conservative equation
\begin{equation}
\label{p1}
\frac{\partial f}{\partial t}=-\frac{\partial Q}{\partial J},
\end{equation}
with a current
\begin{equation}
\label{p2}
Q=-\frac{1}{2}\sum_{m,m'}\int {m A_{mm'}(J,J')^{2}}\delta(m\Omega(J)-m'\Omega(J'))\left\lbrace f(J')m\frac{\partial f}{\partial J}-f(J)m'\frac{\partial f}{\partial J'}\right\rbrace dJ'.
\end{equation}
We impose the boundary conditions $Q=0$ at $J=J_{min}$ and $J=J_{max}$ so as to satisfy the conservation of mass.

\subsection{Equilibrium state} \label{sec_eq}

The Boltzmann distribution
\begin{equation}\label{eq1}
f_{eq}=Ae^{-\beta \epsilon(J)},
\end{equation}
is a stationary solution of the kinetic equation (\ref{q16}). Indeed, using Eq. (\ref{aa6}), we have
\begin{equation}\label{eq2}
\frac{\partial f_{eq}}{\partial J}=-\beta f_{eq}\epsilon'(J)=-\beta f_{eq}\Omega(J). 
\end{equation}
Introducing  this relation in Eq. (\ref{p2}), we see that the integrand is proportional to  
\begin{equation}\label{eq3}
(m\Omega(J)-m'\Omega(J'))\delta (m\Omega(J)-m'\Omega(J'))=0.
\end{equation}
Therefore, the current vanishes and
\begin{equation}\label{eq4}
\frac{\partial f_{eq}}{\partial t}=0.
\end{equation}
An interesting question is whether the Boltzmann distribution is the
{\it only} steady solution of the kinetic equation
(\ref{q16}). Indeed, the current also vanishes if there is no
resonance, i.e. if $m\Omega(J)\neq m'\Omega(J')$ for all orbits
$(m,J)\neq (m',J')$. In the case of point vortices described by a
kinetic equation of the form (\ref{q17}), the Boltzmann distribution
is a steady state but the relaxation stops as soon as the profile of
angular velocity becomes monotonic, so that no resonance is possible,
even if the system is not at statistical equilibrium. This is shown
numerically in
\cite{clkin}. The question is whether a similar situation occurs in
the more complex case of Eq. (\ref{q16}). Note, parenthetically, that
we considered in \cite{curious} a case where there is no resonance. In
that case, other terms must be taken into account in the kinetic
theory and it is found that the system has a curious behavior.

\subsection{Conservation of mass and energy} \label{sec_me}

The conservation of mass  
\begin{equation}\label{me1}
M=\int f(J,t)dJ, 
\end{equation}
is clear from the conservative form of Eq. (\ref{p1}) and the boundary
conditions. Let us show that the kinetic equation (\ref{q16})
conserves the energy
\begin{equation}\label{me2}
E=\int f(J,t)\epsilon(J)dJ.  
\end{equation}
The time derivative of the energy can be written
\begin{equation}\label{me2b}
\dot E=\int \frac{\partial f}{\partial t}(J,t)\epsilon(J)dJ=\int  \Omega(J)Q\; dJ,  
\end{equation}
where we have integrated by parts and used $\epsilon'(J)=\Omega(J)$. Then, introducing Eq. (\ref{p2}) in Eq. (\ref{me2b}), we obtain
\begin{eqnarray}\label{me3}
\dot E=-\frac{1}{2}\sum_{m,m'}\int dJ dJ' \Omega {m A_{mm'}(J,J')^{2}}\delta(m\Omega-m'\Omega')\left\lbrace f'm\frac{\partial f}{\partial J}-f m'\frac{\partial f}{\partial J'}\right\rbrace.
\end{eqnarray}
Interchanging the dummy variables $m$, $m'$ and $J$, $J'$ we find that
\begin{eqnarray}\label{me4}
\dot E=\frac{1}{2}\sum_{m,m'}\int dJ dJ' \Omega' {m' A_{mm'}(J,J')^{2}}\delta(m\Omega-m'\Omega')\left\lbrace f'm\frac{\partial f}{\partial J}-f m'\frac{\partial f}{\partial J'}\right\rbrace,\nonumber\\
\end{eqnarray}
where we  have used that $A_{mm'}(J,J')=A_{m'm}(J',J)$. Summing these two expressions, we obtain
\begin{eqnarray}\label{me5}
\dot E=-\frac{1}{4} \sum_{m,m'}\int dJ dJ' {A_{mm'}(J,J')^{2}}(m\Omega-m'\Omega')\delta(m\Omega-m'\Omega')\left\lbrace f'm\frac{\partial f}{\partial J}-f m'\frac{\partial f}{\partial J'}\right\rbrace,\nonumber\\
\end{eqnarray}
which is equal to zero according to Eq. (\ref{eq3}). Therefore $\dot M=\dot E=0$.

\subsection{H-theorem} \label{sec_h}

Let us show that the entropy 
\begin{equation}\label{h1}
S=-\int f(J,t)\ln f(J,t)dJ,  
\end{equation}
increases. The time derivative of the entropy can be written
\begin{equation}\label{h2}
\dot S=-\int \frac{1}{f}\frac{\partial f}{\partial J} Q dJ.  
\end{equation}
Introducing Eq. (\ref{p2}) in Eq. (\ref{h2}), we obtain
\begin{equation}\label{h3}
\dot S=\frac{1}{2}\int dJ dJ' \frac{1}{f}\frac{\partial f}{\partial J}\sum_{m,m'} {m A_{mm'}(J,J')^{2}}\delta(m\Omega-m'\Omega')\left\lbrace f'm\frac{\partial f}{\partial J}-f m'\frac{\partial f}{\partial J'}\right\rbrace.
\end{equation}
Interchanging the dummy variables $m$, $m'$ and $J$, $J'$ we find that
\begin{equation}\label{h4}
\dot S=-\frac{1}{2}\int dJ dJ' \frac{1}{f'}\frac{\partial f}{\partial J'}\sum_{m,m'} {m' A_{mm'}(J,J')^{2}}\delta(m\Omega-m'\Omega')\left\lbrace f'm\frac{\partial f}{\partial J}-f m'\frac{\partial f}{\partial J'}\right\rbrace.
\end{equation}
Summing these two expressions, we obtain
\begin{equation}\label{h5}
\dot S=\frac{1}{4}\int dJ dJ' \frac{1}{ff'}\sum_{m,m'} {A_{mm'}(J,J')^{2}}\delta(m\Omega-m'\Omega')\left ( f'm\frac{\partial f}{\partial J}-f m'\frac{\partial f}{\partial J'}\right )^{2},
\end{equation}
which is positive. Therefore $\dot S\ge 0$. Note that the above properties persist for the kinetic equation (\ref{q15}).

\section{A more precise kinetic equation} \label{sec_more} 

In the preceding derivation, we have considered that the meanfield
potential was ``frozen''. In fact, this field evolves on a slow
timescale and this evolution must be taken into account in the
advective term of the kinetic equation (\ref{klim4b}). Let us write the
kinetic equation in the form
\begin{equation}
\label{more1}
\frac{\partial f}{\partial t}+\lbrack H,f\rbrack=C(f),
\end{equation} 
where $C(f)$ is the ``collision'' term.  Since the distribution
function (\ref{klim7}) only depends on the energy (approximately), we have
$\lbrack H,f\rbrack\simeq 0$. On the other hand,
\begin{equation}
\label{more2}
\frac{\partial}{\partial t}f(x,v,t)\simeq \frac{\partial f}{\partial t}+\frac{\partial \Phi}{\partial t}\frac{\partial f}{\partial \epsilon},
\end{equation} 
where we have used $\partial\epsilon/\partial t=\partial\Phi/\partial t$ since $\epsilon=v^{2}/2+\Phi(x,t)$. Therefore, the kinetic equation (\ref{more1}) becomes
\begin{equation}
\label{more3}
\frac{\partial f}{\partial t}+\frac{\partial \Phi}{\partial t}\frac{\partial f}{\partial \epsilon}\simeq C(f).
\end{equation} 
The idea now is to average over the orbits in order to eliminate the dependence
on ($x$,$v$), using
\begin{equation}
\label{more4}
\langle X\rangle (\epsilon,t)=\frac{\oint \frac{X(x,v,t)}{v} dx}{\oint \frac{1}{v}dx}, 
\end{equation} 
where we recall that $\oint {dx}/{v}=2\pi g(\epsilon)$ is the density
of states with energy $\epsilon$. The orbit-averaged-kinetic equation
(\ref{more3}) then becomes
\begin{equation}  
\label{more5}
\frac{1}{2\pi}\oint \frac{dx}{v}\left\lbrack \frac{\partial f}{\partial t}+\frac{\partial \Phi} {\partial t}\frac{\partial f}{\partial \epsilon}-C(f)\right\rbrack=0.
\end{equation}
The first term is simply
\begin{equation}
\label{more6}
\frac{1}{2\pi}\oint \frac{dx}{v}\frac{\partial f}{\partial t}=\frac{\partial J}{\partial\epsilon}\frac{\partial f}{\partial t}.
\end{equation}
Noting that
\begin{equation}
\label{more7}
\frac{\partial J}{\partial t}=-\frac{1}{2\pi}\oint \frac{1}{v}\frac{\partial \Phi}{\partial t}dx,
\end{equation}
the second term can be written
\begin{equation}
\label{more8}
\frac{1}{2\pi}\oint \frac{dx}{v}\frac{\partial \Phi} {\partial t}\frac{\partial f}{\partial \epsilon}=-\frac{\partial J}{\partial t}\frac{\partial f}{\partial \epsilon}.
\end{equation}
Finally, the last term is equal to 
\begin{equation}
\label{more9}
\frac{1}{2\pi}\oint \frac{dx}{v} C(f)=\frac{\partial J}{\partial\epsilon}
\frac{\partial Q}{\partial J}=\frac{\partial Q}{\partial \epsilon}.
\end{equation}
Therefore, the final form of the kinetic equation is
\begin{eqnarray}
\label{more10}
\frac{\partial J}{\partial\epsilon}\frac{\partial f}{\partial t}-\frac{\partial J}{\partial t}\frac{\partial f}{\partial \epsilon}
=\frac{1}{2}\frac{\partial}{\partial \epsilon}\sum_{m,m'}\int {m A_{mm'}(J,J')^{2}}\delta(m\Omega(J)-m'\Omega(J'))\nonumber\\
\times \left\lbrace f(J')m\frac{\partial f}{\partial J}-f(J)m'\frac{\partial f}{\partial J'}\right\rbrace dJ',
\end{eqnarray}
where, now, $J=J(\epsilon,t)$.

\section{The thermal bath approximation} \label{sec_tb} 

The kinetic equation (\ref{q16}) describes the evolution of the system
``as a whole''. We have seen that it respects the microcanonical structure
of the initial Hamiltonian system as it conserves mass and
energy. Furthermore, due to the development of correlations between
particles, the Boltzmann entropy increases.  We shall now use this kinetic
theory to describe the relaxation of a test particle in a thermal bath
of field particles. Then, we interprete $f(J,t)=P(J,t)$ in the kinetic
equation (\ref{q16}) as the distribution function of the test particle
and $f(J',t)=f_{0}(J')$ as the {\it static} distribution function of the
field particles with which it interacts. This procedure transforms the
integrodifferential equation (\ref{q16}) into a differential equation
of the Fokker-Planck type \footnote{The Fokker-Planck structure of this equation where the first term in brackets represents a diffusion and the second term a drift (friction) is clear from the discussion of \cite{pa1}, Sec. 2.8. This will also be shown in Sec. \ref{sec_conn}.}:
\begin{equation}
\label{tb1}
\frac{\partial P}{\partial t}=\frac{1}{2}\frac{\partial}{\partial J}\sum_{m,m'}\int {m A_{mm'}(J,J')^{2}}\delta(m\Omega(J)-m'\Omega(J'))\left\lbrace f_{0}(J')m\frac{\partial P}{\partial J}-P(J)m'\frac{d f_{0}}{d J'}\right\rbrace dJ'.
\end{equation}
If we consider that the field particles are in thermal equilibrium,
one has 
\begin{equation}\label{tb2}
f_{0}(J')=Ae^{-\beta \epsilon(J')}.
\end{equation}
Using 
\begin{equation}\label{tb3}
\frac{d f_{0}}{d J'}=-\beta f_{0}(J')\Omega(J'),
\end{equation}
we obtain the Fokker-Planck equation
\begin{equation}
\label{tb4}
\frac{\partial P}{\partial t}=\frac{1}{2}\frac{\partial}{\partial J}\sum_{m,m'}\int {m A_{mm'}(J,J')^{2}}\delta(m\Omega-m'\Omega')f_{0}(J')\left\lbrace  m\frac{\partial P}{\partial J}+\beta P  m'\Omega'\right\rbrace dJ'.
\end{equation}
Using the $\delta$-function, this can be rewritten equivalently
\begin{equation}
\label{tb5}
\frac{\partial P}{\partial t}=\frac{1}{2}\frac{\partial}{\partial J}\sum_{m,m'}\int  {A_{mm'}(J,J')^{2}}\delta(m\Omega-m'\Omega')m^{2} f_{0}(J')\left ( \frac{\partial P}{\partial J}+\beta P \Omega\right ) dJ'.
\end{equation}
Introducing the diffusion coefficient
\begin{equation}
\label{tb6}
D(J)=\frac{1}{2}\sum_{m,m'}\int  {A_{mm'}(J,J')^{2}}\delta(m\Omega-m'\Omega')m^{2} f_{0}(J') dJ',
\end{equation}
we obtain the Fokker-Planck equation
\begin{equation}
\label{tb7}
\frac{\partial P}{\partial t}=\frac{\partial}{\partial J}\left\lbrack D(J)\left (\frac{\partial P}{\partial J}+\beta P \Omega(J)\right )\right\rbrack.
\end{equation}
This equation relaxes towards the Boltzmann distribution
\begin{equation}
\label{tb8}
P_{e}(J)\propto e^{-\beta\epsilon(J)}.
\end{equation}
In terms of the energy, the Fokker-Planck equation (\ref{tb7}) can be rewritten
\begin{equation}
\label{tb9}
\frac{\partial P}{\partial t}=\Omega(\epsilon) \frac{\partial}{\partial \epsilon}\left\lbrack D(\epsilon)\Omega(\epsilon)\left (\frac{\partial P}{\partial \epsilon}+\beta P\right )\right\rbrack.
\end{equation}
Introducing $N(\epsilon,t)=P(\epsilon,t)/\Omega(\epsilon)$, the density of particles with energy $\epsilon$, we obtain
\begin{equation}
\label{tb10}
\frac{\partial N}{\partial t}= \frac{\partial}{\partial \epsilon}\left\lbrack D_{*}(\epsilon)\left (\frac{\partial N}{\partial \epsilon}+\beta N\frac{dV}{d\epsilon}\right )\right\rbrack,
\end{equation}
where 
\begin{equation}
\label{tb11}
D_{*}(\epsilon)=D(\epsilon)\Omega(\epsilon)^{2}=\frac{1}{2}\sum_{m,m'}\int  {A_{mm'}(J,J')^{2}}\delta(m\Omega-m'\Omega')m^{2}\Omega^{2} f_{0}(J') dJ',
\end{equation}
is the diffusion coefficient in energy space and 
\begin{equation}
\label{tb12}
V(\epsilon)=\epsilon+\frac{1}{\beta}\ln\Omega(\epsilon),
\end{equation}
is an effective potential. The stationary solution of Eq. (\ref{tb10}) is
\begin{equation}
\label{tb13}
N_{e}(\epsilon)\propto\frac{1}{\Omega(\epsilon)}e^{-\beta\epsilon}\propto g(\epsilon)e^{-\beta\epsilon},
\end{equation}
corresponding to the Boltzmann distribution. If we now consider a
distribution of the field particles $f_{0}(J)$ such that
$m\Omega(J)\neq m'\Omega(J')$ for any $(m,J)\neq (m',J')$  so
that this distribution is stable on a timescale $Nt_{D}$ (see
Sec. \ref{sec_eq}), the Fokker-Planck equation (\ref{tb1}) becomes
\begin{equation}
\label{addq1}
\frac{\partial P}{\partial t}=\frac{\partial}{\partial J}\left\lbrack D(J)\left (\frac{\partial P}{\partial J}-P\frac{d\ln f_{0}}{dJ}\right )\right\rbrack,
\end{equation}
with a diffusion coefficient
\begin{equation}
\label{addq2}
D(J)=\frac{1}{2}f_{0}(J)\sum_{m}|m|\frac{A_{mm}(J,J)^{2}}{|\Omega'(J)|}.
\end{equation}
To arrive at this expression, we have used the fact that the only
contribution to the collision term comes from $m=m'$ and $J=J'$ and we
have simplified the $\delta$-function according to
$\delta(m(\Omega(J)-\Omega(J')))=\delta(J-J')/|m \Omega'(J)|$.

Note, for comparison, that if we consider the relaxation of a test
vortex in a bath of field vortices with static vorticity distribution
$\omega_{0}(r)$, we obtain from Eq. (\ref{q17}) a Fokker-Planck
equation of the form \cite{preR,kin,clkin}:
\begin{equation}
\label{q17bg}
\frac{\partial P}{\partial t}=-\frac{\gamma}{4r}\frac{\partial}{\partial r}\int_{0}^{+\infty}r'dr'\delta (\Omega(r)-\Omega(r'))\ln\left\lbrack 1-\left (\frac{r_{<}}{r_{>}}\right )^{2}\right\rbrack \left (\frac{1}{r}\omega_{0}'\frac{\partial P}{\partial r}-\frac{1}{r'}P\frac{d \omega_{0}}{d r'}\right ).
\end{equation} 
If we assume that the field vortices are at statistical equilibrium
$\omega_{0}=N\gamma P_{0}=Ae^{-\beta\gamma\psi}$ (thermal bath), the Fokker-Planck equation becomes:
\begin{equation}
\label{tb14}
\frac{\partial P}{\partial t}=\frac{1}{r} \frac{\partial}{\partial r}\left\lbrack rD(r)\left (\frac{\partial P}{\partial r}+\beta \gamma P\frac{d\psi}{dr}\right )\right\rbrack,
\end{equation}
with a diffusion coefficient
\begin{equation}
\label{tb15}
D(r)=-\frac{\gamma}{4r^{2}}\int_{0}^{+\infty}r'\ dr' \omega_{0}(r')\delta(\Omega-\Omega')\ln\left\lbrack 1-\left (\frac{r_{<}}{r_{>}}\right )^{2}\right\rbrack.
\end{equation}
To arrive at these expressions, we have used the $\delta$-function and
the fact that $\Omega(r)=-(1/r)d\psi/dr$. Note that these expressions
are valid even if the profile of angular velocity is non-monotonic
since the Boltzmann distribution is always a stationary solution of
the kinetic equation (\ref{q17}), and it is expected not to evolve at
all (statistical equilibrium state).  If we now consider a
distribution of the field vortices $\omega_{0}(r)$ associated with a
monotonic profile of angular velocity $\Omega(r)$, so that it does not
evolve under the effect of collisions on a timescale $Nt_{D}$, we find that the
relaxation of a test vortex is governed by a Fokker-Planck equation of
the form:
\begin{equation}
\label{tb16}
\frac{\partial P}{\partial t}=\frac{1}{r} \frac{\partial}{\partial r}\left\lbrack rD(r)\left (\frac{\partial P}{\partial r}-P\frac{d\ln\omega_{0}}{dr}\right )\right\rbrack,
\end{equation}
with diffusion coefficient
\begin{equation}
\label{tb15b}
D(r)=\frac{\gamma}{8}\frac{1}{|\Sigma(r)|}\ln N \ \omega_{0}(r),
\end{equation}
where $\Sigma(r)=r\Omega'(r)$ is the local shear created by the field
particles (the angular velocity is related to the vorticity by
$\omega_{0}=(1/r)(r^{2}\Omega)'$). To arrive at this expression, we
have used the fact that the only contribution to the collision term
comes from $r=r'$ and we have simplified the $\delta$-function
according to $\delta(\Omega(r)-\Omega(r'))=\delta(r-r')/|\Omega'(r)|$.
The analogy with the previous equations can be noted.

\section{The homogeneous case} \label{sec_hom} 

In the homogeneous case where $\Phi=0$ and $\epsilon=v^{2}/2$, we have
\begin{equation}
\label{hom1}
\phi=x, \qquad J=v, \qquad \Omega(J)=v.
\end{equation}
On the other hand, 
\begin{equation}
\label{hom2}
A_{mm'}=2\pi\delta_{mm'}\hat{u}_{m},
\end{equation}
where $\hat{u}_{m}$ is the Fourier transform of the
potential. Therefore, the kinetic equation (\ref{q16}) becomes
\begin{equation}
\label{hom3}
\frac{\partial f}{\partial t}=2\pi^{2}\frac{\partial}{\partial J}\sum_{m}\int |m|\hat{u}_{m}^{2}\delta(v-v')\left (f'\frac{\partial f}{\partial v}-f\frac{\partial f}{\partial v'}\right )dv'=0.
\end{equation}
Thus, for a one-dimensional homogeneous system, the collision term
vanishes \cite{bd,cvb,pa1} (this result is well-known in plasma
physics; see last paragraph of \cite{kp}). As a result, the relaxation
time is larger than $Nt_{D}$. This has been shown numerically for the
homogeneous HMF model where it is found that the system relaxes
towards statistical equilibrium on a timescale $N^{1.7}t_{D}$
\cite{yamaguchi}. By contrast, for inhomogeneous systems, the
collision term in Eq. (\ref{q16}) does not necessarily vanish at order
$1/N$ because there are new resonances. Therefore, we may expect that
the system will relax towards statistical equilibrium on a timescale
of order $Nt_{D}$ provided that there are sufficient resonances to
drive the relaxation to completion (see the remarks in
Sec. \ref{sec_eq}).

On the other hand, for a one-dimensional homogeneous system, the
velocity distribution of a test particle evolving in a bath of field
particles with static distribution $f_{0}(v)$ is governed by the
Fokker-Planck equation
\begin{equation}
\label{hom3b}
\frac{\partial P}{\partial t}=2\pi^{2}\frac{\partial}{\partial J}\sum_{m}\int |m|\hat{u}_{m}^{2}\delta(v-v')\left (f_{0}'\frac{\partial P}{\partial v}-P\frac{df_{0}}{d v'}\right )dv'.
\end{equation}
If the field particles are at statistical equilibrium with the
Maxwellian distribution function $f_{0}\propto e^{-\beta v^{2}/2}$ (thermal
bath), the Fokker-Planck equation becomes
\begin{equation}
\label{hom4}
\frac{\partial P}{\partial t}=\frac{\partial}{\partial v}\left\lbrack D(v)\left (\frac{\partial P}{\partial v}+\beta P v\right )\right\rbrack,
\end{equation}
with a diffusion coefficient
\begin{equation}
\label{hom5}
D(v)=2\pi^{2}f_{0}(v)\sum_{m}\hat{u}_{m}^{2}|m|,
\end{equation}
proportional to the distribution function of the thermal bath. In
fact, since the kinetic equation vanishes at order $1/N$, we can
consider a bath with any stable distribution of the Vlasov equation;
it will not change on a timescale $Nt_{D}$. On the other hand,
collective effects can be included in the kinetic theory through the
dielectric function. These generalizations are discussed in
\cite{bd,cvb,pa1}. They lead to a Fokker-Planck equation of the form
\begin{equation}
\label{hom6}
\frac{\partial P}{\partial t}=\frac{\partial}{\partial v}\left\lbrack D(v)\left (\frac{\partial P}{\partial v}-P\frac{d\ln f_{0}}{dv}\right )\right\rbrack,
\end{equation}
with a diffusion coefficient
\begin{equation}
\label{hom7}
D(v)=2\pi^{2}f_{0}(v)\sum_{m}\frac{\hat{u}_{m}^{2}|m|}{\epsilon(m,mv)},
\end{equation}
where  the dielectric function is 
\begin{equation}
\label{hom8}
\epsilon(m,\omega)=1+2\pi\hat{u}_{m}\int \frac{f_{0}'(v)}{\omega/m-v}dv. 
\end{equation}
These kinetic equations are again very similar to the previous equations. For a Maxwellian distribution of the field particles $f_{0}=(\beta/2\pi)^{1/2}\rho\ {\rm exp}(-\beta v^{2}/2)$ (thermal bath), the dielectric function can be explicited \cite{pa1}. The Fokker-Planck equation is then given by Eq. (\ref{hom4}) with a diffusion coefficient
\begin{equation}
\label{hom9}
D(v)=2\pi^{2}f_{0}(v)\sum_{m}\frac{|m|\ \hat{u}_{m}^{2}}{\left\lbrack 1+2\pi\hat{u}_{m} \beta\rho B(\sqrt{\beta\over 2}v)\right \rbrack^{2}+2\pi^{3}\beta^{3}\rho^{2}\hat{u}_{m}^{2}v^{2}e^{-\beta v^{2}}},
\end{equation}
where $B(x)=1-2x e^{-x^{2}}\int_{0}^{x}e^{t^{2}}dt$. The result
(\ref{hom5}) is recovered if we neglect collective effects and take
$\epsilon(m,mv)=1$.

\section{Connection with other works} \label{sec_conn}

While this paper was in course of redaction, we came accross the very
interesting study of Valageas \cite{valageas} for 1D self-gravitating
systems. This author develops a kinetic theory for inhomogeneous
systems by starting directly from the Hamiltonian equations of motion
and using a perturbative expansion of the trajectories in powers of
$1/N$. He can therefore obtain the coefficients of diffusion and
friction directly from the equations of motion. Although the
methodology is different, the expressions of the diffusion and
friction coefficients obtained by Valageas \cite{valageas} are very
similar to those obtained in our approach (we shall see, in fact, that
they coincide after simple transformations). Since this author does
not make the factorization hypothesis (\ref{q7}) this shows that this
assumption can be avoided by using a different approach. We note,
however, that the approach of Valageas does not take into account
collective processes encapsulated in the dielectric function
(polarization cloud). This is precisely in order to obtain this
function that the factorization (\ref{q7}) was necessary in our
approach. Therefore, the agreement between the two approaches when
collective effects are neglected is consistent.

There are, on the
other hand, differences between the two approaches because
Valageas \cite{valageas} finally obtains a Fokker-Planck equation of the form
\begin{equation}
\label{conn1}
\frac{\partial N}{\partial t}= \frac{\partial^{2}}{\partial \epsilon^{2}}(D_{*}N)+\frac{\partial}{\partial \epsilon}(D_{*}\beta N),
\end{equation}
leading to a stationary solution 
\begin{equation}
\label{conn2}
N_{e}(\epsilon)\propto \frac{1}{D_{*}(\epsilon)}e^{-\beta\epsilon},
\end{equation}
which differs from the expected Boltzmann distribution
(\ref{tb13}). In fact, in order to arrive at Eq. (\ref{conn1}),
Valageas makes some approximations that are, in fact, not
necessary. Indeed, he obtains the following expressions for the
diffusion and friction coefficients
\begin{equation}
\label{conn3}
\left \langle \frac{\Delta J}{\Delta t}\right\rangle =\frac{1}{2}\int dJ' f_{0}(J')\sum_{m,m'}\left (m\frac{\partial}{\partial J}-m'\frac{\partial}{\partial J'}\right )A_{mm'}(J,J')^{2}m\delta(m\Omega-m'\Omega'),
\end{equation}
\begin{equation}
\label{conn4}
\left \langle \frac{(\Delta J)^{2}}{\Delta t}\right\rangle =\int dJ' f_{0}(J')\sum_{m,m'}A_{mm'}(J,J')^{2}m^{2}\delta(m\Omega-m'\Omega'),
\end{equation}
where we have used our notations to facilitate the comparison with the
equations of the present paper. If we assume that the dynamics can be
described by a Fokker-Planck equation, the evolution of the
distribution function is given by
\begin{equation}
\label{conn5}
\frac{\partial P}{\partial t}=\frac{1}{2}\frac{\partial^{2}}{\partial J^{2}}\left (\left \langle \frac{(\Delta J)^{2}}{\Delta t}\right\rangle P\right )-\frac{\partial}{\partial J}\left (\left \langle \frac{\Delta J}{\Delta t}\right\rangle P\right ).
\end{equation}
The Fokker-Planck equation (\ref{conn5}) can be rewritten
\begin{equation}
\label{conn6}
\frac{\partial P}{\partial t}=\frac{\partial}{\partial J}\left ( D\frac{\partial P}{\partial J}-P\eta\right ),
\end{equation}
where we have introduced the diffusion coefficient 
\begin{equation}
\label{conn7}
D=\frac{1}{2}\left \langle \frac{(\Delta J)^{2}}{\Delta t}\right\rangle, 
\end{equation}
 and the friction term
\begin{equation}
\label{conn8}
\eta=\left \langle \frac{\Delta J}{\Delta t}\right\rangle-\frac{\partial D}{\partial J}.
\end{equation}
Now, using Eqs. (\ref{conn3}) and (\ref{conn4}), we obtain
\begin{equation}
\label{conn9}
D=\frac{1}{2}\int dJ' f_{0}(J')\sum_{m,m'}A_{mm'}(J,J')^{2}m^{2}\delta(m\Omega-m'\Omega'),
\end{equation}
and
\begin{equation}
\label{conn10}
\eta=-\frac{1}{2}\int dJ' f_{0}(J')\sum_{m,m'}m'\frac{\partial}{\partial J'}A_{mm'}(J,J')^{2}m\delta(m\Omega-m'\Omega').
\end{equation}
Integrating by parts the second equation, we have equivalently
\begin{equation}
\label{conn11}
\eta=\frac{1}{2}\int dJ' \frac{\partial f_{0}}{\partial J'}\sum_{m,m'}A_{mm'}(J,J')^{2}m m' \delta(m\Omega-m'\Omega').
\end{equation}
Therefore, the Fokker-Planck equation (\ref{conn6}) can be rewritten as
\begin{equation}
\label{conn12}
\frac{\partial P}{\partial t}=\frac{1}{2}\frac{\partial}{\partial J}\sum_{m,m'}\int {m A_{mm'}(J,J')^{2}}\delta(m\Omega(J)-m'\Omega(J'))\left\lbrace f_{0}(J')m\frac{\partial P}{\partial J}-P(J)m'\frac{d f_{0}}{d J'}\right\rbrace dJ',
\end{equation}
which coincides with our kinetic equation Eq. (\ref{tb1}). Therefore,
the two theories give the same results and the expressions of the
diffusion and friction terms (\ref{conn3})-(\ref{conn4}) can be
obtained from our approach by starting from Eq. (\ref{tb1}) and using
the steps (\ref{conn3})-(\ref{conn12}) in the other way round as done
in \cite{pa1}.  When we perform a thermal bath approximation as in
Sec. \ref{sec_tb}, we obtain from Eq. (\ref{conn11}) that
$\eta=-D\beta\Omega(J)$ which can be interpreted as an appropriate
Einstein relation. Substituting this relation in Eq. (\ref{conn6}), we
obtain Eqs.  (\ref{tb7}) and (\ref{tb10}) without approximation. These
Fokker-Planck equations relax towards the Boltzmann distribution (\ref{tb13}),
contrary to Eq. (\ref{conn1}) obtained by Valageas
\cite{valageas}. This is more satisfactory on a physical point of
view. Then, if we account for the fact that the distribution of the
bath in Eq. (\ref{conn12}) is not stationary but evolves slowly under
the effect of ``collisions'' in a self-consistent way, we obtain the
integrodifferential kinetic equation (\ref{q16}) or, more precisely,
Eq. (\ref{more10}). Therefore, we can start from the approach of
Valageas (which does not make the factorization hypothesis (\ref{q7}))
to rigorously justify the kinetic equation (\ref{q16}). This equation
was not given by Valageas
\cite{valageas}.  It describes the evolution of the system as a whole
and conserves mass and energy. As we have seen, these conservation
laws and the H-theorem directly result from the symmetrical form of
the kinetic equation (\ref{q16}).

\section{Conclusion} \label{sec_conc}

In this paper, we have introduced a kinetic equation (\ref{q16}) or
(\ref{more10}) that describes the dynamics of spatially inhomogeneous
one-dimensional systems with weak long-range interactions in
angle-action variables. We have shown that this kinetic equation
conserves mass and energy and increases the Boltzmann entropy
(H-theorem). The evolution of the system is due to resonances between
different orbits. The Boltzmann distribution is a stationary solution
of this equation, but it is not necessarily the only one. The system
can be frozen in another steady state of the Vlasov equation if there
is no resonance, i.e. if $m\Omega(J)\neq m'\Omega(J')$ for all orbits
$(m,J)\neq (m',J')$. On the other hand, using a thermal bath
approximation, we have obtained a Fokker-Planck equation (\ref{tb7})
that describes the relaxation of a test particle towards statistical
equilibrium. These nice properties suggest that these kinetic
equations are relevant to describe the evolution of the
system. Unfortunately, the derivation of the kinetic equation
(\ref{q16}) relies on a factorization hypothesis (\ref{q7}) of the
Fourier components of the potential in angle-action variables which is
not rigorously justified. However, we have shown that the recent work
of Valageas \cite{valageas}, which uses a different approach where
this assumption is not required, leads to the same results after some
transformations. Therefore, we can start from this approach to
rigorously derive Eqs. (\ref{q16}), (\ref{more10}) and (\ref{tb7})
which were not given in \cite{valageas}.  Thus, these equations should
provide a good description of the dynamics of inhomogeneous systems
with long-range interactions. These results can be applied to the HMF
model and to the dynamics of point vortices in two dimensions. This
will be presented in a future work
\cite{clkin}. Other extensions are also possible.

\vskip1cm
{\it Acknowledgements}: A preliminary version of this work was
presented to T. Dauxois, F. Bouchet and C. Pichon.  I
also thank C. Sire for useful discussions.

\appendix

\section{The function $A_{mm'}(J,J')$} \label{sec_amn}

The relation between the potential and the distribution function is given by
\begin{equation}
\label{amn1}
\Phi(x)=\int u(x-x')f(x',v')dx'dv'.
\end{equation}
Introducing angle-action variables such that $x=x(\phi,J)$ and
$v=v(\phi,J)$ we obtain
\begin{equation}
\label{amn2}
\Phi(\phi,J)=\int u(x-x')f(\phi',J')d\phi'dJ',
\end{equation}
where we have used $dx dv=d\phi dJ$. Taking the inverse Fourier transform of Eq. (\ref{q2}) and using Eq. (\ref{amn2}), we find that
\begin{equation}
\label{amn3}
\hat \Phi_{m}(J)=\int u(x-x')f(\phi',J')e^{-i m\phi}\frac{d\phi}{2\pi}d\phi'dJ'.
\end{equation}
Inserting Eq. (\ref{q1}) in Eq. (\ref{amn3}) we get
\begin{equation}
\label{amn4}
\hat \Phi_{m}(J)=\sum_{m'}\int u(x-x')\hat{f}_{m'}(J')e^{i(m'\phi'-m\phi)}\frac{d\phi}{2\pi}d\phi'dJ'.
\end{equation}
Defining 
\begin{equation}
\label{amn5}
A_{mm'}(J,J')=\frac{1}{2\pi}\int u(x-x')e^{i(m'\phi'-m\phi)}d\phi d\phi',
\end{equation}
we finally obtain
\begin{equation}
\label{amn6}
\hat \Phi_{m}(J)=\sum_{m'}\int A_{mm'}(J,J') \hat{f}_{m'}(J')dJ'.
\end{equation}

\section{Dispersion relation for the Vlasov equation} \label{sec_dis}

Let us consider the linear dynamical stability of a stationary
solution $f_{0}=f_{0}(J)$ of the Vlasov equation
\begin{equation}\label{dis1}
\frac{\partial f}{\partial t}+\lbrack H,f\rbrack=0.
\end{equation}   
If we substitute
the decomposition $f=f_{0}+\delta f$ in the Vlasov equation (\ref{dis1}) and
neglect nonlinear terms, we find that the perturbation $\delta f$
obeys the linearized Vlasov equation 
\begin{equation}\label{dis2}
\frac{\partial\delta f}{\partial t}+\Omega(J)\frac{\partial\delta f}{\partial\phi}-\frac{\partial\delta\Phi}{\partial\phi}\frac{\partial f}{\partial J}=0.
\end{equation}
The Vlasov and linearized Vlasov equations (\ref{dis1})-(\ref{dis2})
are similar to the Klimontovich and linearized Klimontovich equations
(\ref{klim2})-(\ref{aa13}). It is important to note, however, that in
the Vlasov equation the distribution functions $f$ and $\delta f$ are
{\it smooth} distributions while in the Klimontovich equation they are
expressed in terms of $\delta$-functions.  Writing the perturbation in
the form
\begin{equation}\label{dis3}
\delta f(\phi,J,t)=\sum_{m} \delta\hat{f}_{m}(J,\omega)e^{im\phi}e^{-i\omega t},
\end{equation}
we can proceed exactly like in Sec. \ref{sec_q} except that $g=0$
since the perturbation $\delta f$ of the smooth distribution $f$ does
not involve any $\delta$-function. Therefore, we find that
$\epsilon(\omega)=0$, i.e.
\begin{eqnarray}\label{dis4}
\epsilon(\omega)=1+\sum_{m'}\int A_{m'}(J')^{2}
\frac{m'\frac{\partial f_{0}}{\partial J}(J')}{\omega-m'\Omega(J')}dJ'=0.
\end{eqnarray}
This is the required dispersion relation from which we can study the
linear dynamical stability of the steady state $f_{0}=f_{0}(J)$ with
respect to the Vlasov equation \cite{pichon}. Note that the
integration must be done by using the Landau contour. In general,
$\omega=\omega_{r}+i\omega_{i}$ is a complex pulsation, except at the
point of marginal stability $\omega_{i}=0$ where it is real. In the
kinetic theory of Sec. \ref{sec_q}, we must require that the smooth
meanfield distribution $f(\epsilon)$ is Vlasov stable, which is the
case on the basis of physical grounds (see discussion in
Sec. \ref{sec_klim}). In that case, $\epsilon(\omega_{r})\neq 0$ for
any real $\omega_{r}$.

\end{document}